\documentclass[11pt]{tep}
\pdfoutput=1 
\usepackage{www}

\begin{document}

\thispagestyle{empty}
\def\thefootnote{\fnsymbol{footnote}}
\setcounter{footnote}{1}
\null
\mbox{}\hfill  
FR-PHENO-2017-004
\vskip 0cm
\vfill
\begin{center}
  {\Large \boldmath{\bf 
      Next-to-leading-order QCD and electroweak corrections to\\
      $\PW\PW\PW$ production at proton--proton colliders
    }
    \par} \vskip 2.5em
  {\large
    {\sc Stefan Dittmaier$^{1}$, Alexander Huss$^{2}$ 
      and Gernot Knippen$^{1}$
    }\\[1ex]
    {\normalsize \it 
      $^1$ Albert-Ludwigs-Universit\"at Freiburg, Physikalisches Institut, \\
      D-79104 Freiburg, Germany \\
      $^2$ Institute for Theoretical Physics, ETH, \\
      CH-8093 Z\"urich, Switzerland
    }
    \\[2ex]
  }
  \par \vskip 1em
\end{center}\par
\vskip .0cm \vfill {\bf Abstract:} \par
Triple-W-boson production in proton--proton collisions allows for a direct access to the triple and quartic gauge couplings and provides a window to the mechanism of electroweak symmetry breaking. It is an important process to test the Standard Model (SM) and might be background to physics beyond the SM. We present a calculation of the next-to-leading order (NLO) electroweak corrections to the production of WWW final states at proton--proton colliders with on-shell \PW bosons and combine the electroweak with the NLO QCD corrections. We study the impact of the corrections to the integrated cross sections and to kinematic distributions of the \PW bosons. The electroweak corrections are generically of the size of 5--10\% for integrated cross sections and become more pronounced in specific phase-space regions. The real corrections induced by quark--photon scattering turn out to be as important as electroweak loops and photon bremsstrahlung corrections, but can be reduced by phase-space cuts. Considering that prior determinations of the photon parton distribution function (PDF) involve rather large uncertainties, we compare the results obtained with different photon PDFs and discuss the corresponding uncertainties in the NLO predictions. Moreover, we determine the scale and total PDF uncertainties at the LHC and a possible future $100\,\TeV$ $\Pp\Pp$ collider.
\par
\vskip 1cm
\noindent
May 2017
\par
\null
\setcounter{page}{0}
\clearpage
\def\thefootnote{\arabic{footnote}}
\setcounter{footnote}{0}

\section{Introduction}

After the completion of Run 1 and a successful first phase of Run 2 at the Large Hadron Collider at CERN, many processes predicted by the Standard Model (SM) could be measured and confirmed with an unprecedented precision. However, there are processes that have not been observed so far, but are crucial to our understanding of electroweak (EW) interactions. One such process is the production of three \PW bosons. There is ongoing work directed towards observing this process~\cite{Aaboud:2016ftt} as it represents a great opportunity to experimentally perform a stringent test of the SM. This process allows for a direct handle on the triple and quartic gauge couplings and provides a window to the mechanism of electroweak symmetry breaking in the SM~\cite{Englert:1964et,Higgs:1964pj}.

In order to confront data from colliders with theory and to search for traces of physics beyond the Standard Model (BSM), which may manifest itself in anomalous gauge couplings, precise SM predictions are mandatory. The production of three \PW bosons via proton--proton collision was already calculated at next-to-leading order (NLO) QCD with and without leptonic decays several years ago~\cite{Binoth:2008kt,Campanario:2008yg}. Recently, NLO EW results were published in the narrow width approximation of the \PW bosons~\cite{Yong-Bai:2016sal}. Our NLO calculation of EW and QCD corrections, which is based on on-shell \PW bosons, complements this calculation by presenting additional results and carefully assessing the impact of the uncertainties that arise from the PDFs. The issue of PDF uncertainties is particularly important for $\PW\PW\PW$ production, since quark--photon induced channels have a large impact on the cross section and previous determinations of the photon PDF suffer from large uncertainties. The recently released \propername{LUXqed} photon PDF~\cite{Manohar:2016nzj}, however, is rather precise and stabilizes the prediction considerably.

This paper is structured as follows:
Section~\ref{sec:process} describes the basic properties of $\PW^\pm\PW^\mp\PW^\mp$ production at proton--proton colliders and technical details of our NLO calculation. It further covers the checks and validations we have performed on our calculation. The setup of the calculation and the input parameters are summarized in Sec.~\ref{sec:inputparam}. We present results on total and differential cross sections, determine the scale dependence of the NLO cross section, and assess the error induced by the uncertainty of the PDF in Sec.~\ref{sec:results}. We conclude with Sec.~\ref{sec:conclusion}.
\section{Triple \PW-boson production at NLO}
\label{sec:process}

At leading order (LO), the production of three \PW bosons at proton--proton colliders is induced by the two charge-conjugated partonic subprocesses
\begin{equation}
  u_i \bar{d}_j\ \rightarrow \PWm\PWp\PWp \qquad \text{and} \qquad  \bar{u}_i d_j\  \rightarrow \PWp\PWm\PWm,
\end{equation}
where $i$ and $j$ are the indices of the fermion generation. The different Feynman diagrams contributing to $\PWm\PWp\PWp$ production at LO are shown in Fig.~\ref{fig:lofeynmantop}.
As can be seen from the last diagram, the quartic $\PW\PW\PW\PW$ coupling already enters the LO prediction. The production of three \PW bosons further incorporates Higgs production in association with a \PW boson, where the Higgs particle decays into two \PW bosons. However, owing to the on-shell requirement on the \PW bosons in our calculation, the Higgs boson is restricted to be purely off-shell.

At NLO, additional partons appear in the real emission contributions, namely photons in the NLO EW real emission, gluons in the NLO QCD real emission, and quarks in the gluon-induced and photon-induced channels. A selection of NLO real emission Feynman diagrams is depicted in Fig.~\ref{fig:nlorealphotglugraphs}. In our calculation, infrared (IR) singularities, which arise due to soft and/or collinear emission, are dealt with using the dipole subtraction formalism~\cite{Catani:1996jh,Catani:1996vz,Dittmaier:1999mb,Catani:2002hc,Dittmaier:2008md}. In the virtual contribution, which incorporates additional closed fermion loops and virtual photon, gluon, or weak-vector-boson exchange, we encounter one-loop topologies up to pentagon diagrams. The tensor and scalar loop integrals are evaluated using the \propername{Collier} library~\cite{Denner:2016kdg}, which is based on the results of Refs.~\cite{Denner:2002ii,Denner:2005nn,Denner:2010tr}. Examples of NLO EW loop diagrams are shown in Fig.~\ref{fig:nlovirtgraphs}.
\begin{figure}
  \centering

\begin{tikzpicture}
[line width=0.5pt,
phot/.style={decorate,decoration={snake, segment length=#1, amplitude=2pt}},
phot/.default=6.2pt,
ferm/.style={postaction={decorate},decoration={markings, mark=at position #1 with {\stealtharrow}}},
ferm/.default=0.5]
\small
\coordinate (v1) at (0.8,1.5);
\coordinate (v2) at (0.8,0.75);
\coordinate (v3) at (0.8,0);
\coordinate (i1) at (0,1.5);
\coordinate (i2) at (0,0);
\coordinate (o1) at (1.6,1.5);
\coordinate (o2) at (1.6,0.75);
\coordinate (o3) at (1.6,0);
\draw[phot=6.1pt] (v1) -- (o1);
\draw[phot=6.1pt] (v2) -- (o2);
\draw[phot=6.1pt] (v3) -- (o3);
\draw[ferm] (i1) -- (v1);
\draw[ferm] (v1) -- (v2);
\draw[ferm] (v2) -- (v3);
\draw[ferm] (v3) -- (i2);
\draw[fill=black] (v1) circle (1pt);
\draw[fill=black] (v2) circle (1pt);
\draw[fill=black] (v3) circle (1pt);
\node[xshift=-6pt] at (i1) {$u_i$};
\node[xshift=-6pt] at (i2) {$\bar{d}_j$};
\node[xshift=+10pt] at (o1) {$\PW^+$};
\node[xshift=+10pt] at (o2) {$\PW^-$};
\node[xshift=+10pt] at (o3) {$\PW^+$};
\end{tikzpicture}
\enskip
\begin{tikzpicture}
[line width=0.5pt,
phot/.style={decorate,decoration={snake, segment length=#1, amplitude=2pt}},
phot/.default=6.2pt,
ferm/.style={postaction={decorate},decoration={markings, mark=at position #1 with {\stealtharrow}}},
ferm/.default=0.5]
\small
\coordinate (v1) at (0.4,1.125);
\coordinate (v2) at (1.0,1.125);
\coordinate (v3) at (0.4,0);
\coordinate (i1) at (0,1.125);
\coordinate (i2) at (0,0);
\coordinate (o1) at (1.6,1.5);
\coordinate (o2) at (1.6,0.75);
\coordinate (o3) at (1.6,0);
\draw[phot=6.3pt] (v2) -- (o1);
\draw[phot=6.3pt] (v2) -- (o2);
\draw[phot=6pt] (v3) -- (o3);
\draw[ferm] (i1) -- (v1);
\draw[phot] (v1) -- node[above]{\Pphot/\PZ} (v2);
\draw[ferm] (v1) -- (v3);
\draw[ferm] (v3) -- (i2);
\draw[fill=black] (v1) circle (1pt);
\draw[fill=black] (v2) circle (1pt);
\draw[fill=black] (v3) circle (1pt);
\node[xshift=-6pt] at (i1) {$u_i$};
\node[xshift=-6pt] at (i2) {$\bar{d}_j$};
\node[xshift=+10pt] at (o1) {$\PW^+$};
\node[xshift=+10pt] at (o2) {$\PW^-$};
\node[xshift=+10pt] at (o3) {$\PW^+$};
\end{tikzpicture}
\enskip
\begin{tikzpicture}
[line width=0.5pt,
phot/.style={decorate,decoration={snake, segment length=#1, amplitude=2pt}},
phot/.default=6.2pt,
ferm/.style={postaction={decorate},decoration={markings, mark=at position #1 with {\stealtharrow}}},
ferm/.default=0.5]
\small
\coordinate (v1) at (0.3,0.75);
\coordinate (v2) at (1.0,0.75);
\coordinate (v3) at (1.3,1.125);
\coordinate (i1) at (0,1.5);
\coordinate (i2) at (0,0);
\coordinate (o1) at (1.6,1.5);
\coordinate (o2) at (1.6,0.75);
\coordinate (o3) at (1.6,0);
\draw[phot=6.4pt] (v3) -- (o1);
\draw[phot=6.4pt] (v3) -- (o2);
\draw[phot=6.6pt] (v2) -- (o3);
\draw[ferm] (i1) -- (v1);
\draw[phot=6.1pt] (v1) -- node[below] {\PW} (v2);
\draw[phot] (v2) -- node[above left=-1pt] {\Pphot/\PZ} (v3);
\draw[ferm] (v1) -- (i2);
\draw[fill=black] (v1) circle (1pt);
\draw[fill=black] (v2) circle (1pt);
\draw[fill=black] (v3) circle (1pt);
\node[xshift=-6pt] at (i1) {$u_i$};
\node[xshift=-6pt] at (i2) {$\bar{d}_j$};
\node[xshift=+10pt] at (o1) {$\PW^+$};
\node[xshift=+10pt] at (o2) {$\PW^-$};
\node[xshift=+10pt] at (o3) {$\PW^+$};
\end{tikzpicture}
\enskip
\begin{tikzpicture}
[line width=0.5pt,
phot/.style={decorate,decoration={snake, segment length=#1, amplitude=2pt}},
phot/.default=6.2pt,
ferm/.style={postaction={decorate},decoration={markings, mark=at position #1 with {\stealtharrow}}},
ferm/.default=0.5,
higgs/.style=dashed]
\small
\coordinate (v1) at (0.3,0.75);
\coordinate (v2) at (1.0,0.75);
\coordinate (v3) at (1.3,1.125);
\coordinate (i1) at (0,1.5);
\coordinate (i2) at (0,0);
\coordinate (o1) at (1.6,1.5);
\coordinate (o2) at (1.6,0.75);
\coordinate (o3) at (1.6,0);
\draw[phot=6.4pt] (v3) -- (o1);
\draw[phot=6.4pt] (v3) -- (o2);
\draw[phot=6.6pt] (v2) -- (o3);
\draw[ferm] (i1) -- (v1);
\draw[phot=6.1pt] (v1) -- node[below] {\PW}(v2);
\draw[higgs] (v2) -- node[above left=-1pt] {\PH} (v3);
\draw[ferm] (v1) -- (i2);
\draw[fill=black] (v1) circle (1pt);
\draw[fill=black] (v2) circle (1pt);
\draw[fill=black] (v3) circle (1pt);
\node[xshift=-6pt] at (i1) {$u_i$};
\node[xshift=-6pt] at (i2) {$\bar{d}_j$};
\node[xshift=+10pt] at (o1) {$\PW^+$};
\node[xshift=+10pt] at (o2) {$\PW^-$};
\node[xshift=+10pt] at (o3) {$\PW^+$};
\end{tikzpicture}
\enskip
\begin{tikzpicture}
[line width=0.5pt,
phot/.style={decorate,decoration={snake, segment length=#1, amplitude=2pt}},
phot/.default=6.2pt,
ferm/.style={postaction={decorate},decoration={markings, mark=at position #1 with {\stealtharrow}}},
ferm/.default=0.5]
\small
\coordinate (v1) at (0.4,0.75);
\coordinate (v2) at (1.2,0.75);
\coordinate (i1) at (0,1.5);
\coordinate (i2) at (0,0);
\coordinate (o1) at (1.6,1.5);
\coordinate (o2) at (1.6,0.75);
\coordinate (o3) at (1.6,0);
\draw[phot=6.6pt] (v2) -- (o1);
\draw[phot=7.0pt] (v2) -- (o2);
\draw[phot=6.6pt] (v2) -- (o3);
\draw[ferm] (i1) -- (v1);
\draw[phot=6.1pt] (v1) -- node[above] {\PW} (v2);
\draw[ferm] (v1) -- (i2);
\draw[fill=black] (v1) circle (1pt);
\draw[fill=black] (v2) circle (1pt);
\node[xshift=-6pt] at (i1) {$u_i$};
\node[xshift=-6pt] at (i2) {$\bar{d}_j$};
\node[xshift=+10pt] at (o1) {$\PW^+$};
\node[xshift=+10pt] at (o2) {$\PW^-$};
\node[xshift=+10pt] at (o3) {$\PW^+$};
\end{tikzpicture}
  \caption{Feynman diagrams of the process $\Pp\Pp\rightarrow\PW^-\PW^+\PW^++X$ at LO. The indices $i$, $j$ mark the fermion generation.}
  \label{fig:lofeynmantop}
  \vspace*{\abovecaptionskip}
\begin{tikzpicture}
[line width=0.5pt,
phot/.style={decorate,decoration={snake, segment length=#1, amplitude=2pt}},
phot/.default=6.2pt,
ferm/.style={postaction={decorate},decoration={markings, mark=at position .5 with {\stealtharrow}}},
higgs/.style=dashed]
\small
\coordinate (v1) at (0.5,1.05);
\coordinate (v2) at (1.2,1.05);
\coordinate (v3) at (2.0,1.75);
\coordinate (v4) at (2.0,0.35);
\coordinate (i1) at (0,1.75);
\coordinate (i2) at (0,0.35);
\coordinate (o1) at (2.5,2.1);
\coordinate (o2) at (2.5,1.4);
\coordinate (o3) at (2.5,0.7);
\coordinate (o4) at (2.5,0);
\node[xshift=-5pt] at (i1) {$\Pqu$};
\node[xshift=-5pt] at (i2) {$\Paqd$};
\node[xshift=+10pt] at (o1) {$\PWp$};
\node[xshift=+10pt] at (o2) {$\PWm$};
\node[xshift=+10pt] at (o3) {$\PWp$};
\node[xshift=+7pt] at (o4) {$\Pphot$};
\draw[ferm] (i1) -- (v1);
\draw[ferm] (v1) -- (i2);
\draw[phot] (v1) -- node[pos=0.6,above]{$\PW$} (v2);
\draw[higgs] (v2) -- node[pos=0.4,above=2pt]{$\PH$} (v3);
\draw[phot] (v2) -- node[pos=0.3,below=6pt]{$\PW$} (v4);
\draw[phot=6.5pt] (v3) -- (o1);
\draw[phot=6.5pt] (v3) -- (o2);
\draw[phot=6.5pt] (v4) -- (o3);
\draw[phot=6.5pt] (v4) -- (o4);
\draw[fill=black] (v1) circle (1pt);
\draw[fill=black] (v2) circle (1pt);
\draw[fill=black] (v3) circle (1pt);
\draw[fill=black] (v4) circle (1pt);
\end{tikzpicture}
\enskip
\begin{tikzpicture}
[line width=0.5pt,
phot/.style={decorate,decoration={snake, segment length=#1, amplitude=2pt}},
phot/.default=6.2pt,
glu/.style={decorate,decoration={coil, aspect=0.9, segment length=#1, amplitude=2.5pt}},
glu/.default=5pt,
ferm/.style={postaction={decorate},decoration={markings, mark=at position .5 with {\stealtharrow}}},
higgs/.style=dashed]
\small
\coordinate (v1) at (0.9,2.0);
\coordinate (v2) at (0.9,1.05);
\coordinate (v3) at (0.9,0.1);
\coordinate (v4) at (1.8,1.05);
\coordinate (i1) at (0,2.1);
\coordinate (i2) at (0,0);
\coordinate (o1) at (2.3,2.1);
\coordinate (o2) at (2.3,1.4);
\coordinate (o3) at (2.3,0.7);
\coordinate (o4) at (2.3,0);
\node[xshift=-5pt] at (i1) {$\Pqu$};
\node[xshift=-5pt] at (i2) {$\Paqd$};
\node[xshift=+10pt] at (o1) {$\PWp$};
\node[xshift=+10pt] at (o2) {$\PWm$};
\node[xshift=+10pt] at (o3) {$\PWp$};
\node[xshift=+7pt] at (o4) {$\Pg$};
\draw[ferm] (i1) -- (v1);
\draw[ferm] (v1) -- node[left=2pt]{$\Pqd$} (v2);
\draw[ferm] (v2) -- node[left=2pt]{$\Pqd$} (v3);
\draw[ferm] (v3) -- (i2);
\draw[phot=6.5pt] (v1) -- (o1);
\draw[glu=6.4pt] (v3) -- (o4);
\draw[phot=6.45pt] (v2) -- node[above]{$\PZ$}(v4);
\draw[phot=6.6pt] (v4) -- (o2);
\draw[phot=6.6pt] (v4) -- (o3);
\draw[fill=black] (v1) circle (1pt);
\draw[fill=black] (v2) circle (1pt);
\draw[fill=black] (v3) circle (1pt);
\draw[fill=black] (v4) circle (1pt);
\end{tikzpicture}%
  \enskip%
\begin{tikzpicture}
[line width=0.5pt,
phot/.style={decorate,decoration={snake, segment length=#1, amplitude=2pt}},
phot/.default=6.2pt,
glu/.style={decorate,decoration={coil, aspect=0.9, segment length=#1, amplitude=2.5pt}},
glu/.default=5pt,
ferm/.style={postaction={decorate},decoration={markings, mark=at position .5 with {\stealtharrow}}},
higgs/.style=dashed]
\small
\coordinate (v1) at (0.9,1.4);
\coordinate (v2) at (0.9,0.1);
\coordinate (v3) at (1.8,1.4);
\coordinate (i1) at (0,1.5);
\coordinate (i2) at (0,0);
\coordinate (o1) at (2.5,2.1);
\coordinate (o2) at (2.5,1.4);
\coordinate (o3) at (2.5,0.7);
\coordinate (o4) at (2.5,0);
\node[xshift=-5pt] at (i1) {$\Pqu$};
\node[xshift=-5pt] at (i2) {$\gamma$};
\node[xshift=+10pt] at (o1) {$\PWp$};
\node[xshift=+10pt] at (o2) {$\PWm$};
\node[xshift=+10pt] at (o3) {$\PWp$};
\node[xshift=+7pt] at (o4) {$\Pqd$};
\draw[ferm] (i1) -- (v1);
\draw[ferm] (v1) -- node[left=2pt]{$\Pqd$} (v2);
\draw[phot] (i2) -- (v2);
\draw[ferm] (v2) -- (o4);
\draw[phot=6.2pt] (v1) -- node[above]{$\PW$}(v3);
\draw[phot=6.7pt] (v3) -- (o1);
\draw[phot=6.2pt] (v3) -- (o2);
\draw[phot=6.7pt] (v3) -- (o3);
\draw[fill=black] (v1) circle (1pt);
\draw[fill=black] (v2) circle (1pt);
\draw[fill=black] (v3) circle (1pt);
\end{tikzpicture}
\enskip
\begin{tikzpicture}
[line width=0.5pt,
phot/.style={decorate,decoration={snake, segment length=#1, amplitude=2pt}},
phot/.default=6.2pt,
glu/.style={decorate,decoration={coil, aspect=0.9, segment length=#1, amplitude=2.5pt}},
glu/.default=5pt,
ferm/.style={postaction={decorate},decoration={markings, mark=at position .5 with {\stealtharrow}}},
higgs/.style=dashed]
\small
\coordinate (v1) at (0.5,1.05);
\coordinate (v2) at (1.2,1.05);
\coordinate (v3) at (1.6,0.7);
\coordinate (v4) at (2.0,0.35);
\coordinate (i1) at (0,1.75);
\coordinate (i2) at (0,0.35);
\coordinate (o1) at (2.4,2.1);
\coordinate (o2) at (2.4,1.4);
\coordinate (o3) at (2.4,0.7);
\coordinate (o4) at (2.4,0);
\node[xshift=-5pt] at (i1) {$\Paqd$};
\node[xshift=-5pt] at (i2) {$\Pg$};
\node[xshift=+10pt] at (o1) {$\PWp$};
\node[xshift=+10pt] at (o2) {$\PWm$};
\node[xshift=+10pt] at (o3) {$\PWp$};
\node[xshift=+7pt] at (o4) {$\Paqu$};
\draw[ferm] (v1) -- (i1);
\draw[glu=6.3pt] (i2) -- (v1);
\draw[ferm] (v2) -- node[above]{$\Pqd$} (v1);
\draw[ferm] (v3) -- node[pos=0.2,left=2.5pt]{$\Pqu$} (v2);
\draw[ferm] (v4) -- node[pos=0.2,left=2.5pt]{$\Pqd$} (v3);
\draw[phot=6.3pt] (v2) -- (o1);
\draw[phot=6.5pt] (v3) -- (o2);
\draw[phot=5.7pt] (v4) -- (o3);
\draw[ferm] (o4) -- (v4);
\draw[fill=black] (v1) circle (1pt);
\draw[fill=black] (v2) circle (1pt);
\draw[fill=black] (v3) circle (1pt);
\draw[fill=black] (v4) circle (1pt);
\end{tikzpicture}
  \caption{Example diagrams contributing to the NLO real emission (from left to right: photonic emission, gluonic emission,  quark--photon, and quark--gluon induced channels) of the process $\Pp\Pp\rightarrow\PWm\PWp\PWp+X$.}
  \label{fig:nlorealphotglugraphs}
  \vspace*{\abovecaptionskip}

\begin{tikzpicture}
[line width=0.5pt,
phot/.style={decorate,decoration={snake, segment length=#1, amplitude=2pt}},
phot/.default=6.2pt,
ferm/.style={postaction={decorate},decoration={markings, mark=at position .5 with {\stealtharrow}}}]
\small
\coordinate (v1) at (0.8,1.5);
\coordinate (v2) at (2.1,1.5);
\coordinate (v3) at (2.1,0.3);
\coordinate (v4) at (0.8,0.3);
\coordinate (v5) at (0.8,0.9);
\coordinate (i1) at (0,1.8);
\coordinate (i2) at (0,0);
\coordinate (o1) at (2.7,1.8);
\coordinate (o2) at (2.7,0.9);
\coordinate (o3) at (2.7,0.0);
\node[xshift=-5pt] at (i1) {$\Pqu$};
\node[xshift=-5pt] at (i2) {$\Paqd$};
\node[xshift=+10pt] at (o1) {$\PWp$};
\node[xshift=+10pt] at (o2) {$\PWm$};
\node[xshift=+10pt] at (o3) {$\PWp$};
\draw[ferm] (i1) -- (v1);
\draw[ferm] (v1) -- node[left=2pt]{$\Pqd$} (v5);
\draw[ferm] (v5) -- node[left=2pt]{$\Pqu$} (v4);
\draw[ferm] (v4) -- (i2);
\draw[phot] (v1) -- node[above]{$\PW$}(v2);
\draw[dashed] (v2) -- node[pos=0.2,right]{$\PH$}(v3);
\draw[phot] (v3) -- node[below]{$\PW$}(v4);
\draw[phot=6.1pt] (v2) -- (o1);
\draw[phot] (v5) -- (o2);
\draw[phot=6.1pt] (v3) -- (o3);
\draw[fill=black] (v1) circle (1pt);
\draw[fill=black] (v2) circle (1pt);
\draw[fill=black] (v3) circle (1pt);
\draw[fill=black] (v4) circle (1pt);
\draw[fill=black] (v5) circle (1pt);
\end{tikzpicture}
\qquad
\begin{tikzpicture}
[line width=0.5pt,
phot/.style={decorate,decoration={snake, segment length=#1, amplitude=2pt}},
phot/.default=6.2pt,
ferm/.style={postaction={decorate},decoration={markings, mark=at position .5 with {\stealtharrow}}}]
\small
\coordinate (v1) at (0.5,1.4);
\coordinate (v2) at (0.5,0.4);
\coordinate (v3) at (1.2,1.4);
\coordinate (v4) at (1.8,1.4);
\coordinate (v5) at (2.5,1.4);
\coordinate (i1) at (0,1.8);
\coordinate (i2) at (0,0);
\coordinate (o1) at (2.8,1.8);
\coordinate (o2) at (2.8,0.9);
\coordinate (o3) at (2.8,0);
\node[xshift=-5pt] at (i1) {$\Pqu$};
\node[xshift=-5pt] at (i2) {$\Paqd$};
\node[xshift=+10pt] at (o1) {$\PWp$};
\node[xshift=+10pt] at (o2) {$\PWm$};
\node[xshift=+10pt] at (o3) {$\PWp$};
\draw[ferm] (i1) -- (v1);
\draw[ferm] (v1) -- node[left=2pt]{$\Pqu$} (v2);
\draw[ferm] (v2) -- (i2);
\draw[phot] (v1) -- node[above]{$\gamma$} (v3);
\draw[ferm] (v3) to[controls=+(90:0.4) and +(90:0.4)] node[above]{$f$} (v4);
\draw[ferm] (v4) to[controls=+(-90:0.4) and +(-90:0.4)] node[below]{$f$} (v3);
\draw[phot] (v4) -- node[above]{$\PZ$} (v5);
\draw[phot] (v5) -- (o1);
\draw[phot] (v5) -- (o2);
\draw[phot=6.25pt] (v2) -- (o3);
\draw[fill=black] (v1) circle (1pt);
\draw[fill=black] (v2) circle (1pt);
\draw[fill=black] (v3) circle (1pt);
\draw[fill=black] (v4) circle (1pt);
\draw[fill=black] (v5) circle (1pt);
\end{tikzpicture}
\qquad
\begin{tikzpicture}
[line width=0.5pt,
phot/.style={decorate,decoration={snake, segment length=#1, amplitude=2pt}},
phot/.default=6.2pt,
glu/.style={decorate,decoration={coil, aspect=0.9, segment length=#1, amplitude=2.5pt}},
glu/.default=5pt,
ferm/.style={postaction={decorate},decoration={markings, mark=at position .5 with {\stealtharrow}}},
higgs/.style=dashed,
cross/.style={cross out, draw, minimum size=2*(#1-\pgflinewidth), inner sep=0pt, outer sep=0pt, line width = 1.5pt},
cross/.default=4pt]
\small
\coordinate (v1) at (0.5,1.3);
\coordinate (v2) at (0.5,0.5);
\coordinate (v3) at (1.5,1.3);
\coordinate (v4) at (1.5,0.5);
\coordinate (v5) at (2.2,0.5);
\coordinate (i1) at (0,1.8);
\coordinate (i2) at (0,0);
\coordinate (o1) at (2.7,1.8);
\coordinate (o2) at (2.7,0.9);
\coordinate (o3) at (2.7,0.0);
\node[xshift=-5pt] at (i1) {$\Pqu$};
\node[xshift=-5pt] at (i2) {$\Paqd$};
\node[xshift=+10pt] at (o1) {$\PWp$};
\node[xshift=+10pt] at (o2) {$\PWm$};
\node[xshift=+10pt] at (o3) {$\PWp$};
\draw[ferm] (i1) -- (v1);
\draw[ferm] (v1) -- node[above]{$\Pqu$} (v3);
\draw[ferm] (v2) -- (i2);
\draw[ferm] (v4) -- node[below]{$\Pqd$}(v2);
\draw[glu] (v1) -- node[right=2pt]{$\Pg$}(v2);
\draw[ferm] (v3) -- node[right]{$\Pqd$} (v4);
\draw[phot] (v4) -- node[below]{$\gamma$} (v5);
\draw[phot=6.0pt] (v3) -- (o1);
\draw[phot=6.7pt] (v5) -- (o2);
\draw[phot=6.32pt] (v5) -- (o3);
\draw[fill=black] (v1) circle (1pt);
\draw[fill=black] (v2) circle (1pt);
\draw[fill=black] (v3) circle (1pt);
\draw[fill=black] (v4) circle (1pt);
\draw[fill=black] (v5) circle (1pt);
\end{tikzpicture}
  \caption{Selection of one-loop Feynman diagrams contributing to the NLO virtual corrections for the process $\Pp\Pp\rightarrow\PWm\PWp\PWp+X$.}
  \label{fig:nlovirtgraphs}
\end{figure}

We have performed two completely independent calculations: one where the amplitudes are generated and further processed with the packages \propername{FeynArts}~\cite{Hahn:2000kx} and \propername{FormCalc}~\cite{Hahn:1998yk}, and a second calculation using in-house software based on Feynman diagrams generated with \propername{FeynArts 1}~\cite{Kublbeck:1990xc}. The results of the two calculations agree within the Monte Carlo integration errors. We have further checked that the results do not depend on the regularization scheme, employing either mass or dimensional regularization for the treatment of IR divergences.

To check for consistency, we also compare with NLO results on $\PW\PW\PW$ production available in the literature~\cite{Binoth:2008kt,Campanario:2008yg,Yong-Bai:2016sal}: Tables~\ref{tab:checkqcd} and~\ref{tab:checknlo} show the comparison of our results to the results in the literature and agreement is found within the Monte Carlo integration errors, with the exception of the QCD result of Ref.~\cite{Yong-Bai:2016sal}, where a phenomenologically insignificant deviation is found at the few per mill level.
\begin{table}
  \centering
  \caption{Comparison to published results on NLO QCD corrections to the total cross section for $\PWm\PWp\PWp$ production at the LHC and a center-of-mass (CM) energy of $\sqrt{s}=14\,\TeV$. The input parameter were chosen as reported in Ref.~\cite{Binoth:2008kt} with $\mur=\muf=3\MW$. Contributions from associated Higgs production are omitted here.}
  \label{tab:checkqcd}
  \begin{tabular}{cd{2.5}d{3.5}}
    \toprule
    Reference & \ccol{$\sigma^\LO$ {\small[fb]}} & \ccol{$\sigma^{\NLO\,\QCD}$ {\small[fb]}}\\
    \midrule
    our results & 82.725(11) & 145.25(3)\\
   \cite{Binoth:2008kt} & 82.5(5) & 146.2(6) \\
   \cite{Campanario:2008yg} & 82.8(1) & 145.2(3) \\ 
   \cite{Yong-Bai:2016sal} & 82.74(3) & 145.17(6) \\ 
    \bottomrule
  \end{tabular}
\end{table}
\begin{table}
  \centering
  \caption{Comparison to published results on NLO correction to the total cross section for $\PWm\PWp\PWp$ production at the LHC and the CM energy of $\sqrt{s}=14\,\TeV$. Input parameters and the definition of the NLO cross section and the relative corrections were chosen as reported in Ref.~\cite{Yong-Bai:2016sal} with $\mur=\muf=\nicefrac{3}{2}\MW$.}
  \label{tab:checknlo}
  \begin{tabular}{cd{3.7}d{3.5}d{3.5}d{2.6}d{2.5}}
    \toprule
    Reference & \ccol{$\sigma^\LO$ {\small{[fb]}}} & \ccol{$\sigma^\NLO$ {\small{[fb]}}} & \ccol{$\delta^\QCD$ {\small{[\%]}}} & \ccol{$\delta^\EW_{\Pq\Paq}$ {\small{[\%]}}} & \ccol{$\delta^\EW_{\Pq\gamma}$ {\small{[fb]}}}\\
    \midrule
    our results & 78.645(10) & 186.42(6) & 106.96(4) & -4.199(5) & 18.73(2)\\
    \cite{Yong-Bai:2016sal} & 78.65(1) & 187.04(9) & 107.50 & -4.16 & 18.77\\
    \bottomrule
  \end{tabular}
\end{table}
\section{Input parameter scheme}
\label{sec:inputparam}

We follow the recent Yellow Report of the LHC Higgs Cross Section Group \cite{deFlorian:2016spz} and adopt the following input parameters,
\begin{equation}
  \begin{aligned}
    \MW&=80.385\,\GeV,\ &  \MZ&=91.1876\,\GeV, \qquad\qquad \MH=125\,\GeV,\\
    m_\Pqb&=m_\Pqb^{\OS}=4.92\,\GeV, &  m_\Pqt&=m_\Pqt^{\OS}=172.5\,\GeV,\\
    \alphas(\MZ)&=0.118, &\Gmu &=1.1663787\times10^{-5}\,\GeV^{-2},
  \end{aligned}
\end{equation}
where the superscript OS stands for the on-shell scheme.
We neglect all fermion masses except for bottom- and top-quark masses, and further ignore the negligible mixing involving the third generation quarks.%
\footnote{Note that due to the negligible mixing involving quarks of the third generation, the bottom quark never occurs as an external state in our calculation but only appears in closed fermion loops.
The inclusion of the bottom-quark mass in this case is an improvement to a massless treatment.}
As a result, the CKM matrix factorizes from all matrix elements and can be absorbed into the parton luminosities.%
\footnote{
Owing to the mass degeneracy among the quarks of the first two generations and the absence of mixing with the third generation in our setup, the dependence on the CKM matrix drops out whenever a summation over internal and external final-state flavors is performed. 
The only case where the unitarity of the CKM matrix cannot be exploited in this way is when the up- and down-type quark are both in the initial state and thus receive different weights from the PDFs. 
The calculation can still be performed using a diagonal CKM by absorbing the CKM factors into the parton luminosities in this case.}
Furthermore, the SM behaves like a CP-conserving theory in our calculation. We calculate with a block-diagonal CKM matrix, where the mixing among the first two generations is parametrized by the Cabibbo angle $\theta_\text{Cabibbo}=0.22731$, so that the relevant CKM entries are given by
\begin{equation}
  \left\lvert V_{\Pqu\Pqd}\right\rvert=\left\lvert V_{\Pqc\Pqs}\right\rvert=0.97428,\qquad\left\lvert V_{\Pqu\Pqs}\right\rvert=\left\lvert V_{\Pqc\Pqd}\right\rvert=0.22536.
\end{equation}

We work in the \Gmu-scheme (see, e.g., Ref.~\cite{Dittmaier:2001ay}), where the electromagnetic coupling $\alpha=\alphagmu$ is a derived quantity and given by
\begin{equation}
\alphagmu=\frac{\sqrt{2}}{\pi}\Gmu\MW^2\left(1-\frac{\MW^2}{\MZ^2}\right).
\end{equation}
The \Gmu-scheme accounts for universal corrections to the $\rho$ parameter and the running of $\alpha$ from the Thomson limit to the electroweak scale. The running of the strong coupling constant \alphas is taken from the PDF set used.

We employ a dynamic renormalization and factorization scale (\mur and \muf, respectively) given by
\begin{equation}
  \mur=\muf=\sqrt{\Big(3\MW\Big)^2+\Big(\sum\nolimits_{i\in S}\mathbf{p}_{\text{T},i}\Big)^2},
  \label{eq:centralscale}
\end{equation}
where $\mathbf{p}_{\text{T},i}$ is the (vectorial) transverse momentum of particle $i$ and $S$ is the set of all outgoing particles which carry no color. 
Note that this scale choice is equal to the production threshold $3\MW$ if there are no color-charged particles in the final state.
In order to estimate the residual theory uncertainties from missing higher-order corrections, we examine the scale dependence in Sect.~\ref{sec:results} by varying the scales with respect to the central choice~\eqref{eq:centralscale} by factors of $\nicefrac{1}{2}$ and $2$.

We use the \propername{LHAPDF6} library \cite{Buckley:2014ana} to perform the convolution of the partonic cross sections with the PDFs. We calculate the pure LO cross section, denoted by $\sigma^\LO$, with the LO \propername{NNPDF3.0} set~\cite{Ball:2014uwa}. All NLO contributions, including the LO contribution to the NLO cross section $\sigma^\LO_1$, are calculated with the NLO \propername{NNPDF3.0QED}~\cite{Ball:2014uwa,Ball:2013hta} set. The photon-induced contributions are calculated with the \propername{LUXqed} set~\cite{Manohar:2016nzj}. Since all PDFs in the \propername{LUXqed} set, except for the photon PDF, are based on the \propername{PDF4LHC} NNLO set \cite{Butterworth:2015oua}, the error introduced by using different PDFs for the quark--photon-induced and every other channel should be negligible in the overall PDF uncertainty. We additionally provide results using the \propername{NNPDF3.0QED} and the \propername{CT14QED} (inclusive)~\cite{Schmidt:2015zda} PDF sets in the quark--photon-induced channels to better assess the corresponding uncertainty. 
Throughout this work we use PDF sets with five active flavors.

\section{Numerical results}
\label{sec:results}

In the following we present our results in terms of relative corrections defined as
\begin{equation}
  \deltaqqew \equiv \frac{\Delta\sigma^{\NLO\,\EW}_{\Pq\Paq^\prime}}{\sigma^\LO_1}, \qquad 
  \deltaqaew \equiv \frac{\Delta\sigma^{\NLO\,\EW}_{\Pq\Pphot}}{\sigma^\LO}, \qquad 
  \deltaqcd \equiv \frac{\sigma^{\LO}_1+\Delta\sigma^{\NLO\,\QCD}}{\sigma^\LO}-1,
  \label{eq:delta}
\end{equation}
where the subscripts $\Pq\Paq^\prime$ and $\Pq\Pphot$ indicate the partonic channels. We combine the QCD corrections and the EW corrections to the quark-induced channels multiplicatively and include the photon-induced correction in an additive manner, so that the total NLO cross section is given by
\begin{equation}
\sigma^\NLO=\left[\left(1+\deltaqqew\right)\left(1+\deltaqcd\right)+\deltaqaew\right]\times\sigma^\LO.
\end{equation}
This approach is well motivated by the factorization of EW logarithms, which dominate the EW correction $\deltaqqew$ at large energies, from the long-range QCD effects and is preferable over a purely additive approach~\cite{Chiu:2008vv,Chiu:2009mg,Manohar:2014vxa}.
Note that by normalizing the QCD correction to the LO cross section evaluated with LO PDFs, the term $1+\deltaqcd$ is identical to the usual definition of the $K$-factor, up to small QED corrections in the PDFs. Moreover, the EW correction factor $\deltaqqew$ becomes rather insensitive to the PDF choice and the factorization scale.

\begin{table}
  \centering
  \caption{LO and full NLO cross sections, $\sigma^\LO$ and $\sigma^\NLO$, as well as the NLO relative corrections, $\delta$, at different CM energies $\sqrt{s}$ of the collider. The indicated errors are estimates for the Monte Carlo integration errors.}
  \label{tab:xs}
  \begin{subtable}{\linewidth}
    \centering
    \caption{$\Pp\Pp\rightarrow\PWm\PWp\PWp+X$}
    \begin{tabular}{d{3}d{1.9}d{1.10}d{1.2}d{2.2}d{3.2}}
      \toprule
      \ccol{$\sqrt{s}$ {\small[TeV]}} & \ccol{$\sigma^\LO$ {\small[pb]}} & \ccol{$\sigma^\NLO$ {\small[pb]}} & \ccol{\deltaqqew\ {\small[\%]}}  & \ccol{\deltaqaew\ {\small[\%]}} & \ccol{\deltaqcd\ {\small[\%]}}\\
      \midrule
      7 & 0.029407(3) & 0.044217(5) & -3.35 & 5.67 & 49.70\\
      8 & 0.037090(4) & 0.057237(7) & -3.52 & 6.59 & 53.12\\
      13 & 0.079476(11) & 0.13587(3) & -4.09 & 10.71 & 67.08\\
      14 & 0.088496(12) & 0.15375(2) & -4.17 & 11.46 & 69.34\\
      100 & 0.98056(16) & 2.6574(4) & -5.40 & 41.30 & 142.84\\
      \bottomrule
    \end{tabular}
  \end{subtable}\\
  \vspace*{1em}
  \begin{subtable}{\linewidth}
    \centering
    \caption{$\Pp\Pp\rightarrow\PWp\PWm\PWm+X$}
    \begin{tabular}{d{3}d{1.11}d{1.10}d{1.2}d{2.2}d{3.2}}
      \toprule
      \ccol{$\sqrt{s}$ {\small[TeV]}} & \ccol{$\sigma^\LO$ {\small[pb]}} & \ccol{$\sigma^\NLO$ {\small[pb]}} & \ccol{\deltaqqew\ {\small[\%]}}  & \ccol{\deltaqaew\ {\small[\%]}} & \ccol{\deltaqcd\ {\small[\%]}}\\
      \midrule
      7 & 0.0139183(15) & 0.021580(2) & -3.00 & 6.40 & 53.23\\
      8 & 0.018136(2) & 0.028865(3) & -3.16 & 7.34 & 56.77\\
      13 & 0.043278(5) & 0.076369(12) & -3.69 & 11.58 & 71.20\\
      14 & 0.048927(6) & 0.087752(14) & -3.77 & 12.36 & 73.53\\
      100 & 0.72097(11) & 1.9987(4) & -5.08 & 42.01 & 147.81\\
      \bottomrule
    \end{tabular}
  \end{subtable}
\end{table}
Table~\ref{tab:xs} shows the LO and full NLO cross sections, as well as the relative corrections defined in Eq.~\eqref{eq:delta}. We consider different LHC and possible high-energy proton--proton collider energies for the two $\PW\PW\PW$ final states. The relative corrections for $\PWm\PWp\PWp$ and $\PWp\PWm\PWm$ are very similar. The NLO corrections are dominated by the QCD corrections which amount to a $K$-factor of $\sim1.7$ at LHC energies of $13\,\TeV$ and $14\,\TeV$. The photon-induced contributions are positive and overcompensate the negative EW corrections of the quark-induced channels, leading to total EW corrections of $\sim6\%$ and $\sim7\%$ at the current LHC energy of $13\,\TeV$ for the two charge-conjugated final states, respectively. Note that this partial cancellation is not systematic in the sense that the two compensating effects are not directly correlated. Due to the impact of the EW corrections it is important to take the EW corrections into account, when comparing data to theory. We estimate the missing higher-order EW corrections to be of the order of the squares of the individual NLO corrections, i.e.\ $\sim1\%$ for LHC energies. We observe that the EW corrections in the pure quark-induced channels, which are generically of $\order{\sim5\%}$, show only very little sensitivity to the collider energy. The quark--photon-induced contribution, on the other hand, rises with the $\Pp\Pp$ scattering energy, reaching $\sim40\%$ for the scenario of a future $100\,\TeV$ collider. This demonstrates the importance of determining the photon PDF precisely for high-energy proton--proton scattering. The QCD corrections increase with growing collider CM energy owing to the higher gluon luminosity. The large $K$-factors of $\sim1.7$ ($\sim2.5$) at $13\,\TeV$ ($100\,\TeV$), which are driven by the quark--gluon-induced channels, ask for further improvements by higher-order QCD corrections. At least improvements by multi-jet merging seem mandatory.

\subsection{Differential distributions}

\begin{figure}
  \centering
  \includegraphics[trim=0 0.3cm 0 0.45cm]{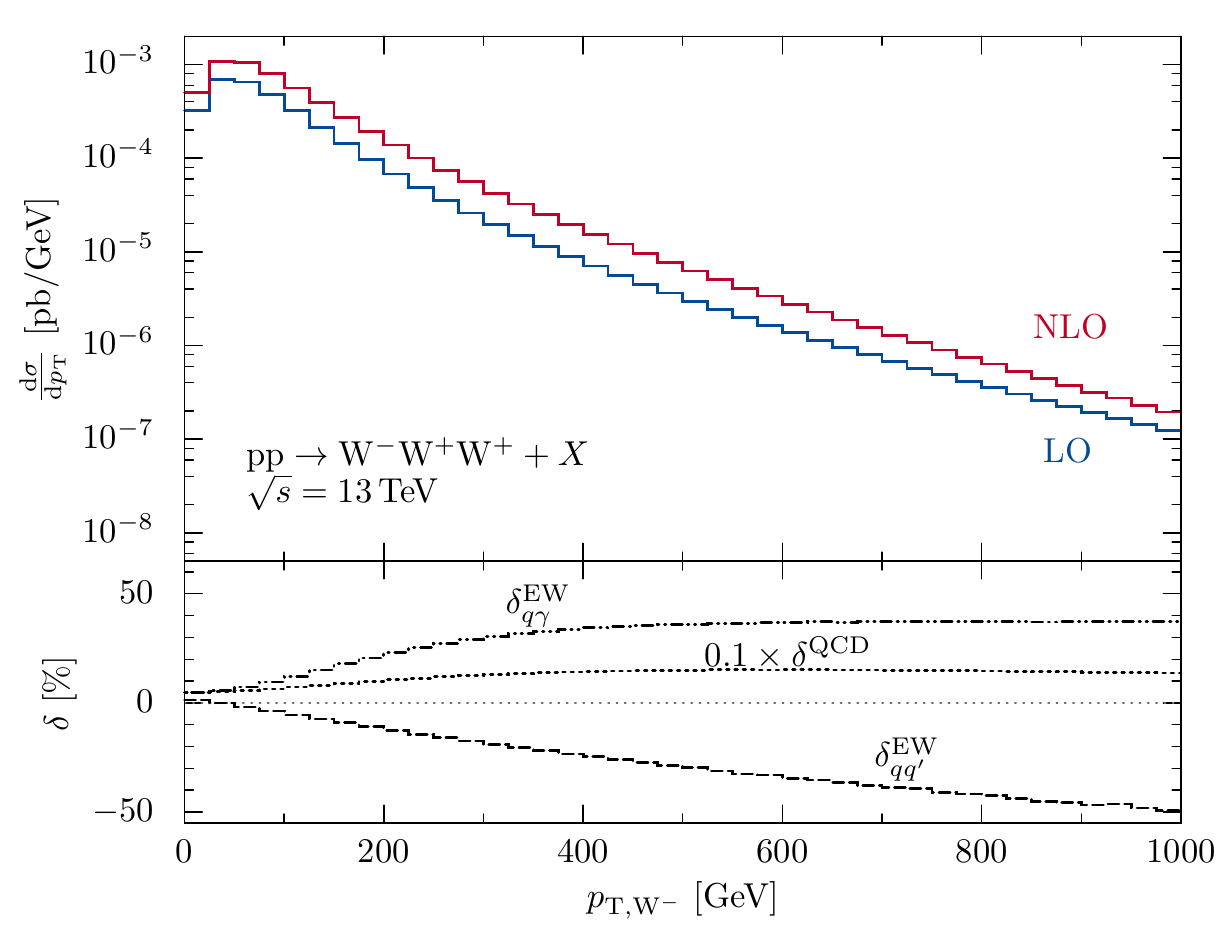}
  \caption{Transverse-momentum distribution of the distinct negatively charged \PW boson in $\PWm\PWp\PWp$ production. The lower panel shows the size of the different relative corrections. The curve of the QCD correction is scaled down by a factor of $0.1$.}
  \label{fig:dsptWm}
\end{figure}

\begin{figure}
  \centering
  \includegraphics[trim=0 0.3cm 0 0.45cm]{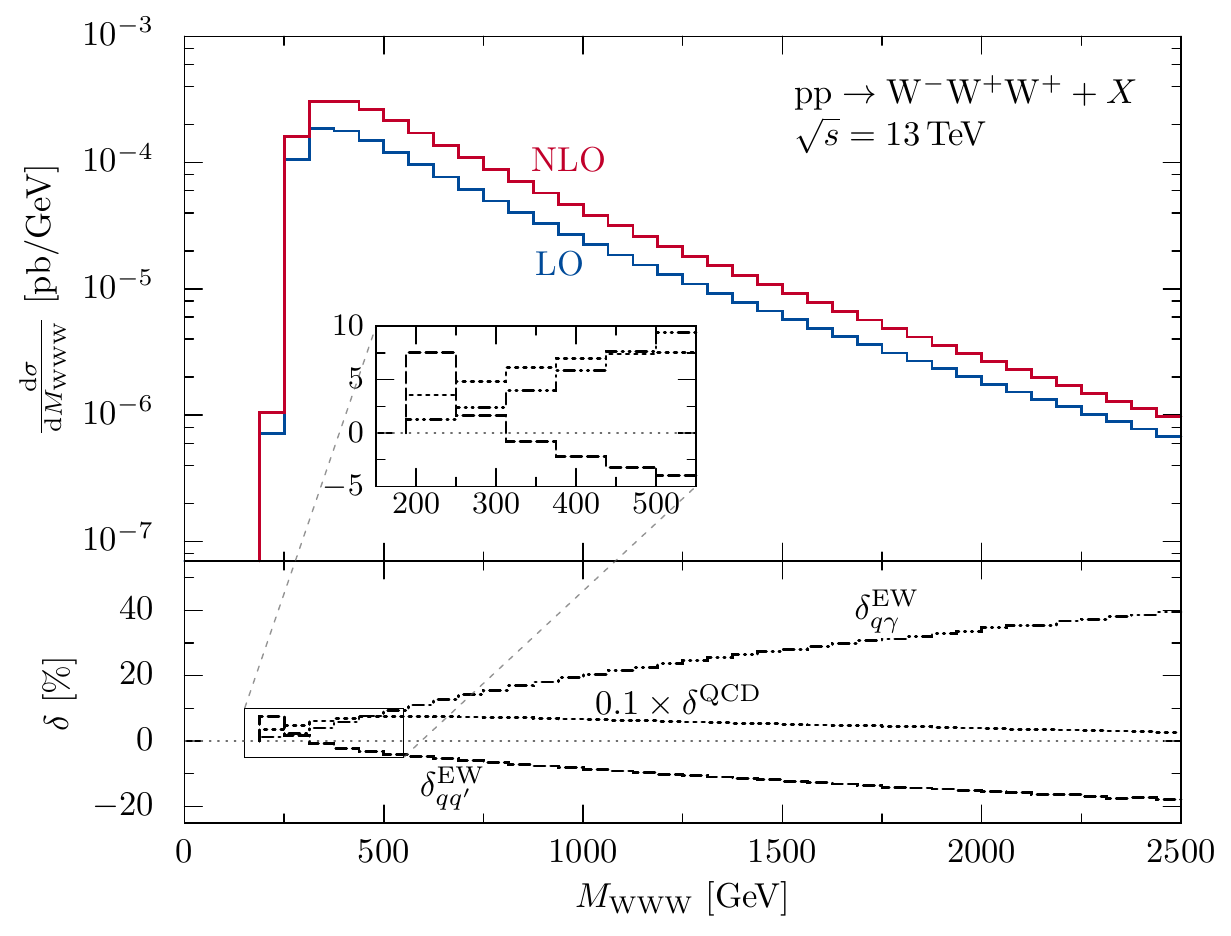}
  \caption{Differential cross section over the total invariant mass of the $\PW\PW\PW$ system. The lower panel shows the size of the different relative corrections. Details on the relative corrections for small invariant masses can be seen in the magnified cutout. The curve of the QCD correction is scaled down by a factor of $0.1$.}
  \label{fig:dsMWWW}
\end{figure}

\begin{figure}
  \centering
  \includegraphics[trim=0 0.3cm 0 0.45cm]{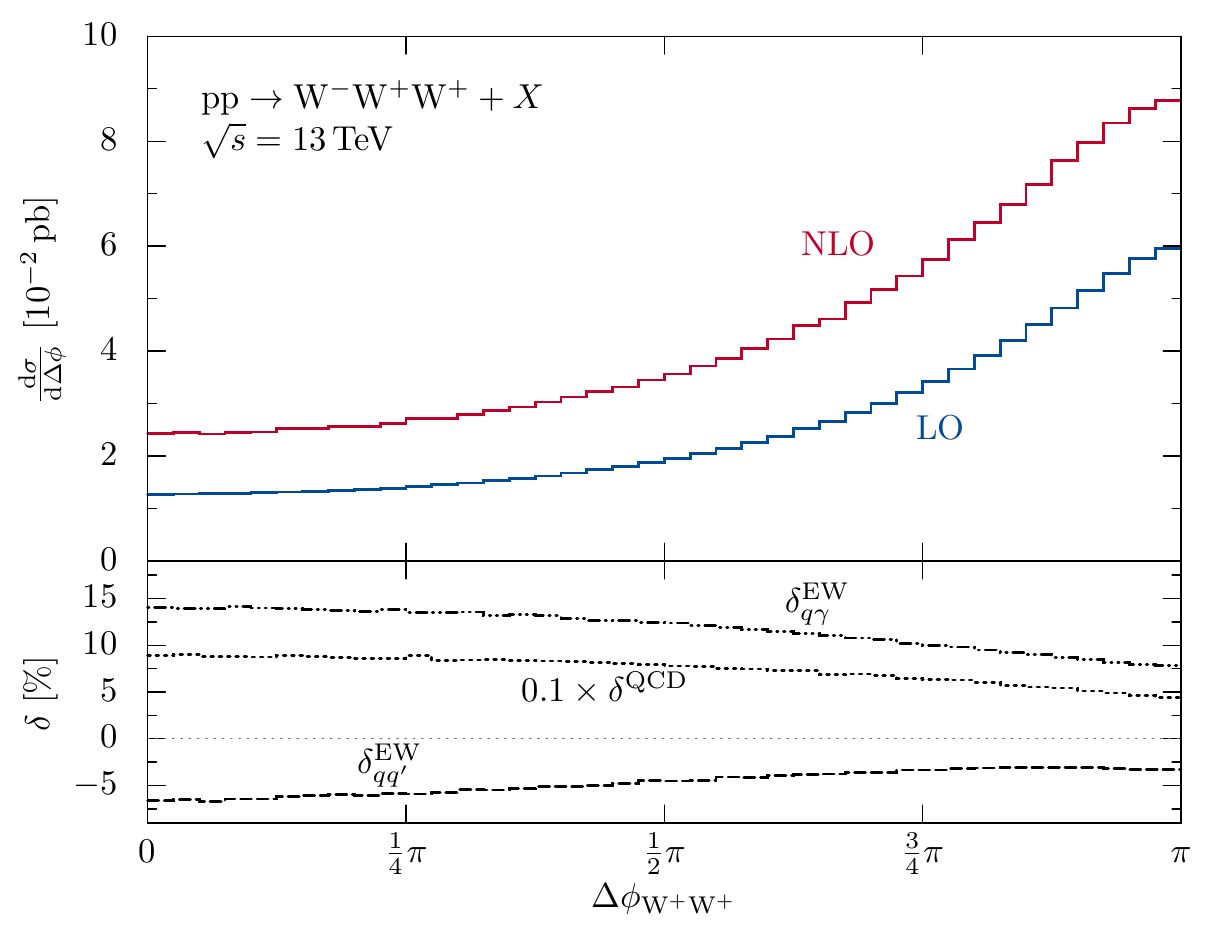}
  \caption{Distribution in the difference of the azimuthal angles of the two positively charged \PW bosons in $\PWm\PWp\PWp$ production. The lower panel shows the size of the different relative corrections. The curve of the QCD correction is scaled down by a factor of $0.1$.}
  \label{fig:dsdphiWpWp}
\end{figure}

Due to the valence quark content of the protons, $\PWm\PWp\PWp$ production is the dominant production mode among the two charge-conjugated processes. As both final states can be easily separated, we will focus on the dominant, positively charged final state in the following. A selection of differential distributions including a breakdown of the corrections into the relative factors defined in Eq.~\eqref{eq:delta} is presented in Figs.~\ref{fig:dsptWm}, \ref{fig:dsMWWW}, and~\ref{fig:dsdphiWpWp}. Note that the size of these corrections will be inherited by distributions based on decay leptons when dropping the on-shell requirement on the \PW bosons. 
As high transverse momenta of the \PWm or high total invariant masses correspond to high partonic CM energies, the unitarizing effect of Higgs exchanges can be seen in the drop of the associated differential distributions shown in Figs.~\ref{fig:dsptWm} and~\ref{fig:dsMWWW}. In this high-energy regime, Sudakov logarithms from soft EW gauge-boson exchange are the leading contribution to the EW correction in the quark--induced channel, yielding corrections of several $-10\,\%$ which can even overrule the large quark--photon-induced corrections at very high $p_\mathrm{T}$. At low invariant masses near the production threshold, the effect of the Coulomb singularity, which arises due to photon exchange between \PW bosons, is visible. In this region the leading behavior of the NLO EW correction $\deltaqqew$ is dominated by
\begin{equation*}
\delta_\text{Coul}\sim\pm\frac{\alpha\pi}{2\beta_\PW},
\end{equation*}
where $\beta_\PW$ is the velocity of the \PW bosons of any $\PW^+\PW^\mp$ boson pair in their (two-particle) CM frame~\cite{Sommerfeld:1939}. Even though the QCD corrections grow with increasing $p_\mathrm{T}$ of the \PW boson they are rather independent on the total invariant mass. Figure~\ref{fig:dsdphiWpWp} shows that the NLO QCD correction changes the shape of the distribution in the difference of the azimuthal angle, preferring smaller angle differences. This effect is slightly enhanced by the total EW correction.

\subsection{NLO $\PW\PW\PW$ cross sections with a jet veto}

\begin{figure}
  \centering
  \includegraphics[trim=0 0.2cm 0 0.4cm]{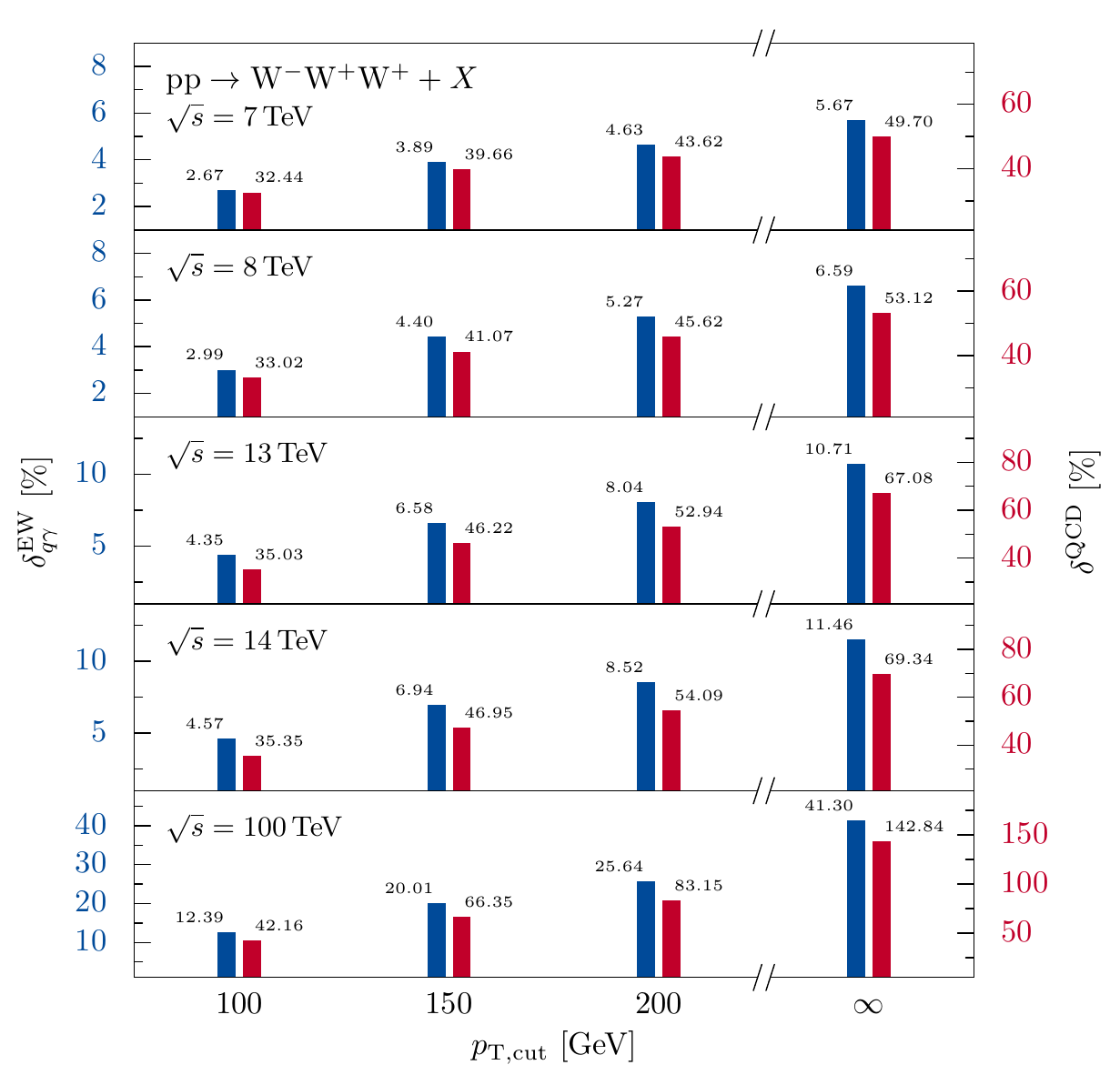}
  \caption{Relative correction in the quark--photon-induced channel (red) and relative NLO QCD correction (blue) for different cut values $p_{\mathrm{T},\text{cut}}$ and collider energies for the process $\Pp\Pp\rightarrow\PWm\PWp\PWp+X$. The values at $p_{\mathrm{T},\text{cut}}=\infty$ are the relative corrections of the full integrated cross section without any jet veto.}
  \label{fig:relativecorrectionjetveto}
\end{figure}

The large impact of the quark--photon-induced channel on the total cross section can be reduced by restricting the phase space of the additional jet in the final state by a jet veto. To this end, we require the transverse momentum of the additional outgoing parton, which can be experimentally identified with a jet, to be below a certain threshold value $p_{\mathrm{T},\text{cut}}$. This threshold should not be chosen too small in order to not affect the effective cancellation of IR singularities. 
Otherwise, large logarithms of the jet-veto cut would remain in the final result requiring resummation~\cite{Banfi:2012jm}, which however is beyond the scope of this work.
As we cut on the transverse momentum of the jet alone, only the quark--photon-induced channel, the quark--gluon-induced channel and the QCD real emission contribution are affected. The integrated cross sections for different values of $p_{\mathrm{T},\text{cut}}$ are presented in Tab.~\ref{tab:xsjetveto}. In Fig.~\ref{fig:relativecorrectionjetveto} the impact of the $p_\mathrm{T}$-cut on the relative corrections in $\PWm\PWp\PWp$ production is shown. A relatively strong cut at a transverse momentum of $100\,\GeV$ reduces the total NLO cross section by $\sim 23\%$ at the current LHC CM energy of $13\,\TeV$. In detail, the QCD correction drops by a factor of $\sim2$ and the photon-induced channel decreases to approximately $40\%$ of its original value. With increasing CM energy the impact of the $p_\mathrm{T}$-cut increases. In combination with the strong growth of the quark--photon-induced and the QCD corrections (see above) this results in a reduction of the NLO cross section by $\sim50\%$ for a value of $p_{\mathrm{T},\text{cut}}=100\,\GeV$.
\begin{table}
  \centering
  \caption{NLO cross sections $\sigma^\NLO$ [pb] with phase-space cut on the (leading) jet transverse momentum at different CM energies $\sqrt{s}$ of the collider. The error is an estimated integration error resulting from the Monte Carlo integration.}
  \label{tab:xsjetveto}
  \begin{subtable}{\linewidth}
    \centering
    \caption{$\Pp\Pp\rightarrow\PWm\PWp\PWp+X$}
    \begin{tabular}{d{3}d{1.9}d{1.9}d{1.9}d{1.9}}
      \toprule
      \ccol{$\sqrt{s}$ {\small[TeV]}} & \ccol{$p_\mathrm{T}(\text{jet}) {<} 100\,\GeV$} & \ccol{$p_\mathrm{T}(\text{jet}) {<} 150\,\GeV$} & \ccol{$p_\mathrm{T}(\text{jet}) {<} 200\,\GeV$} & \ccol{no cut} \\
      \midrule
      7 & 0.038428(5) & 0.040837(5) & 0.042181(5) & 0.044217(5)\\
      8 & 0.048708(8) & 0.052112(7) & 0.054066(8) & 0.057237(7)\\
      13 & 0.10638(2) & 0.11669(2) & 0.12297(2) & 0.13587(3)\\
      14 & 0.11883(2) & 0.13076(3) & 0.13822(2) & 0.15375(2)\\
      100 & 1.4403(6) & 1.7396(6) & 1.9505(6) & 2.6574(4)\\
      \bottomrule
    \end{tabular}
  \end{subtable}\\
  \vspace{1em}
  \begin{subtable}{\linewidth}
    \centering
    \caption{$\Pp\Pp\rightarrow\PWp\PWm\PWm+X$}
    \begin{tabular}{d{3}d{1.10}d{1.9}d{1.9}d{1.9}}
      \toprule
      \ccol{$\sqrt{s}$ {\small[TeV]}} & \ccol{$p_\mathrm{T}(\text{jet}) {<} 100\,\GeV$} & \ccol{$p_\mathrm{T}(\text{jet}) {<} 150\,\GeV$} & \ccol{$p_\mathrm{T}(\text{jet}) {<} 200\,\GeV$} & \ccol{no cut}\\
      \midrule
      7 & 0.018748(2) & 0.019961(2) & 0.020626(2) & 0.021580(2)\\
      8 & 0.024576(3) & 0.026338(3) & 0.027339(3) & 0.028865(3)\\
      13 & 0.059995(11) & 0.065900(12) & 0.069501(12) & 0.076389(12)\\
      14 & 0.068036(17) & 0.075005(14) & 0.079316(13) & 0.087752(14)\\
      100 & 1.0926(4) & 1.3215(3) & 1.4824(5) & 1.9987(4)\\
      \bottomrule
    \end{tabular}
  \end{subtable}
\end{table}
\subsection{Discussion of PDF uncertainties in the photon-induced channel}
\label{sec:photonpdferror}

The inclusion of QED corrections into the determination of PDFs was first considered by the MRST collaboration~\cite{Martin:2004dh}, which imposed strong model assumptions on the parametrization of the photon PDF and based the fit mostly on DIS data. Later, the NNPDF~\cite{Ball:2013hta} and CT~\cite{Schmidt:2015zda} collaborations provided photon PDFs as well, the former without any model assumptions and using mostly LHC data to constrain the photon PDF, the latter with similar, but less strict assumptions than the MRST collaboration and using $\Pe\Pp\rightarrow\Pe\Pphot+X$ data. Recently, a new approach was put forward which made it possible to derive the photon PDF directly from the proton structure functions $F_2(x,Q^2)$ and $F_L(x,Q^2)$, which are well determined from $\Pe\Pp$ scattering data. This new \propername{LUXqed}~\cite{Manohar:2016nzj} PDF set exhibits a very small uncertainty. Another approach, using the structure functions as well, was used by Harland-Lang et al.~\cite{Harland-Lang:2016apc}. Not long ago, the xFitter collaboration published results on a photon PDF fit to high-mass Drell--Yan data at the LHC~\cite{Giuli:2017oii}.

Until recently, the photon PDF was largely unconstrained due to the limited amount of data. Therefore, the uncertainty on quark--photon and photon--photon-induced contributions could be easily as large as the contributions themselves. The quark--photon-induced channel of $\PW\PW\PW$ production constitutes the largest contribution to the NLO EW corrections. Thus, an uncertainty estimate is essential for a meaningful physical prediction.

\begin{figure}
  \centering
  \begin{minipage}[t]{.4\linewidth}
    \vspace{0pt}
    \begin{tabular}{cd{1.6}}
      \toprule
      PDF set & \ccol{$\deltaqaew$ [\%]}\\
      \midrule
      \propername{MRSTQED04} & \ccol{13.99}\\
      \propername{NNPDF3.0QED} & 6.88^{+9.96}_{-1.03}\\
      \propername{CT14QEDinc} & 10.87^{+1.40}_{-1.40}\\
      \propername{LUXqed} & 10.71^{+0.08}_{-0.08}\\
      \bottomrule
    \end{tabular}
  \end{minipage}
  \begin{minipage}[t]{.57\linewidth}
    \vspace{0pt}
    \includegraphics[trim=0 0.15cm 0.15cm 0.3cm]{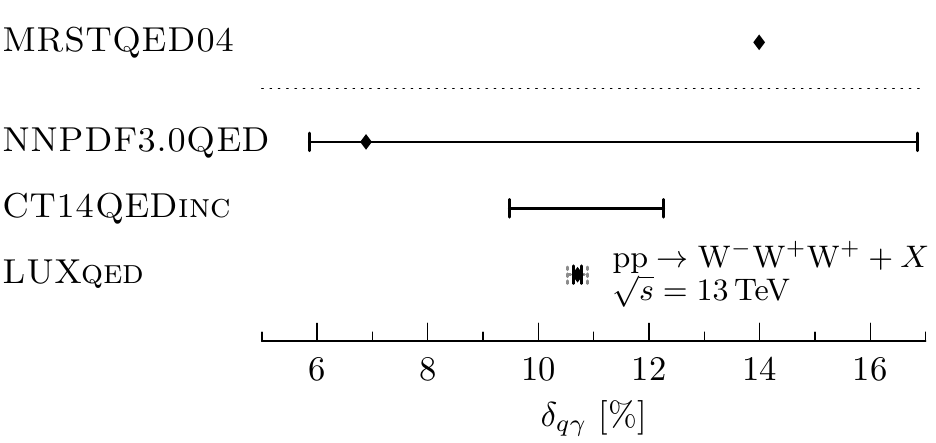}
  \end{minipage}
  \caption{Quark--photon-induced correction and uncertainties due to the photon PDF for the different PDF sets including QED corrections for the production of $\PWm\PWp\PWp$ at a CM energy of 13\,\TeV. A central value for the CT14QEDinc PDF set was calculated by taking the midpoint of the range of cross sections calculated in the table on the left. The additional dashed gray error bar for the \propername{LUXqed} PDF set shows the total PDF uncertainty for the quark--photon-induced correction.}
  \label{fig:photonerrorcorrection}
\end{figure}

\begin{figure}
  \centering
  \includegraphics[trim=0 0.1cm 0 0.45cm]{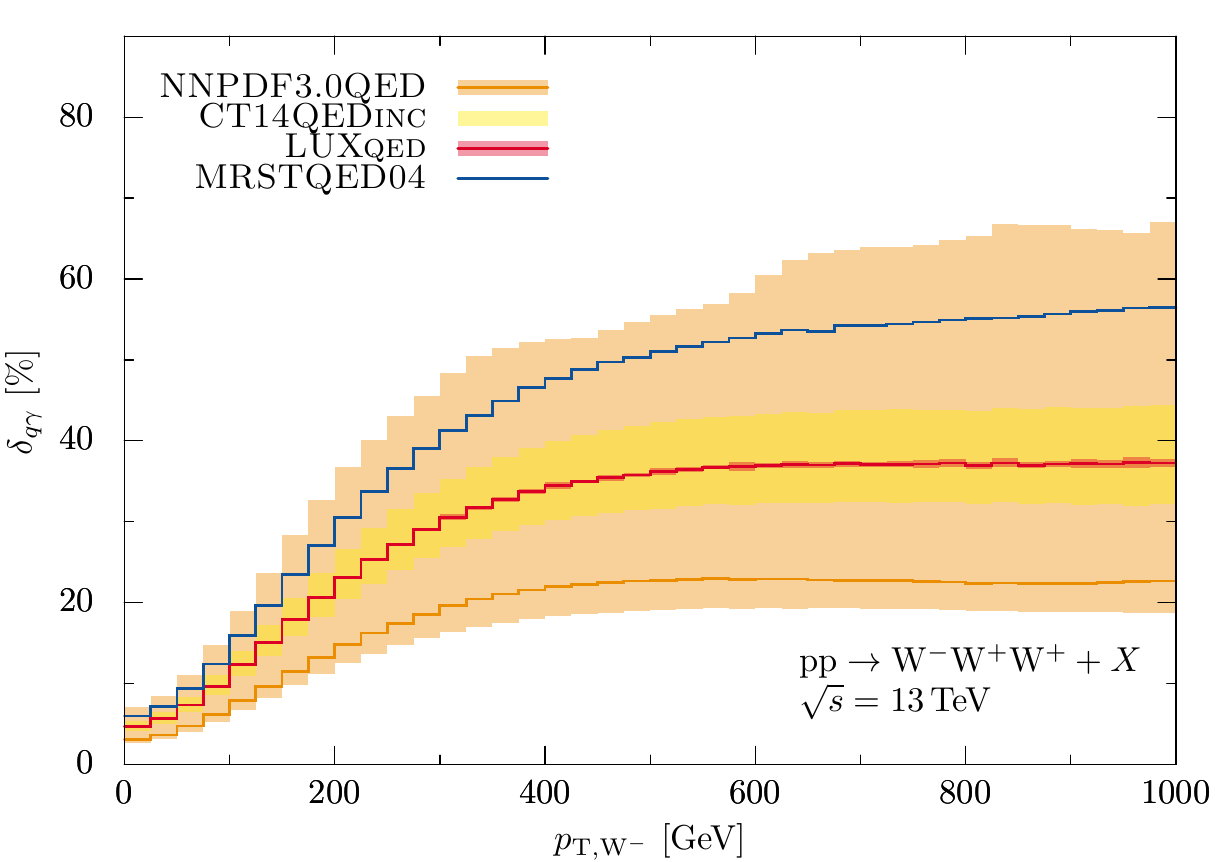}
  \caption{Differential distribution of the quark--photon-induced correction over the transverse momentum of \PWm with uncertainties due to the photon PDF for different PDF sets incorporating QED corrections.}
  \label{fig:photonerrordistribution}
\end{figure}

We assess the uncertainty by analyzing our results with different available PDF sets incorporating QED corrections: \propername{NNPDF3.0QED}, \propername{CT14QEDinc}, and \propername{LUXqed}. We include numbers for the largely outdated \propername{MRSTQED04} set as well, but do not use it in the uncertainty estimate. The uncertainty of the photon PDF of the NNPDF set is highly non-Gaussian and only loosely constrained by data. Following the procedure used by the LHC Higgs Cross Section Working Group~\cite{deFlorian:2016spz} in the calculation of the Higgs production cross sections via vector-boson fusion and Higgs-strahlung, we therefore take the lower limit of the cross sections calculated separately with all \propername{NNPDF3.0QED} replicas as the lower bound, the median of the cross sections as the central value, and the maximum of the 68\% smallest cross sections as the upper bound. Note that obtaining an error estimate based on the uncertainty of the photon PDF alone is difficult in the case of the \propername{NNPDF3.0QED} sets as there are no dedicated variations for the photon PDF. We have computed the \propername{NNPDF3.0QED} errors both using the full PDF error for the quark--photon channel and fixing the incoming quark to the central member PDF and only varying the photon PDF over the replicas. Both approaches result in similar error estimates. The \propername{CT14QEDinc} PDF set does not give any information on a central value, only a range of their free fit parameter, the initial photon momentum fraction at the fit scale $p_0^\gamma$. At the 68\% confidence level,  $p_0^\gamma$ is restricted to be between 0\% and 0.11\%. This range yields the error bar used. The error on the \propername{LUXqed} PDF set was calculated as described in the corresponding paper~\cite{Manohar:2016nzj}. As the \propername{LUXqed} PDF set uses the Hessian method with symmetric eigenvectors to describe the uncertainties, the variance of a cross section $\sigma$ is given by
\begin{equation}
  \var(\sigma) = \sum\limits_{j=1}^{N_\text{EV}} \left(\sigma_j-\sigma_0\right)^2,
\end{equation}
where $N_\text{EV}$ is the number of eigenvector PDF sets, $\sigma_j$ the cross section evaluated with eigenvector set $j$ and $\sigma_0$ the central value. The \propername{MRSTQED04} set does not provide any uncertainty information. Figure~\ref{fig:photonerrorcorrection} shows the central values of the photon-induced contribution, where the error bars represent the photon-PDF uncertainty for the different PDF sets. In Fig.~\ref{fig:photonerrordistribution} the impact of the photon PDF uncertainty on differential distributions is shown.
We observe that the results based on the recent PDF sets are consistent with each other. Due to the limited amount of data and no model assumptions, the uncertainty of the \propername{NNPDF3.0QED} set is the largest. The photon PDF of the \propername{LUXqed} PDF set shows an outstanding small uncertainty, which is even less than the remaining PDF uncertainties (c.f.\ Fig.~\ref{fig:photonerrorcorrection}).

\subsection{Estimating the uncertainties of the total cross sections}

We investigate the two main sources to the uncertainty of the total cross sections: missing higher-order corrections estimated through the residual dependence on the factorization and renormalization scale, and the uncertainties of the PDFs.

At LO, the production of three \PW bosons at a proton--proton collider is a purely electroweak process. As a consequence, there is no dependence on the renormalization scale $\muR$ at LO, so that no reduction of the scale dependence when going from LO to NLO is expected. In order to estimate the scale uncertainties we vary our scale choice in Eq.~\eqref{eq:centralscale} up and down by a factor of 2. The scale uncertainty of the total LO and NLO cross section for different collider CM energies is shown in Tab.~\ref{tab:nloscalepdf}.

Another significant contribution to the total uncertainty is the overall PDF uncertainty which is given by the square root of the variance defined by
\begin{equation}
  \begin{split}
    \left.\var\left(\sigma^\NLO\right)\right|_\text{PDF}= &\left.\var\left(\left(1+\deltaqqew\right)\left(1+\deltaqcd\right)\sigma^\LO\right)\right|_\text{PDF} + \left.\var\left(\Deltasigmaqaew\right)\right|_\text{PDF}\\
    & + \left.2\cov\left(\left(1+\deltaqqew\right)\left(1+\deltaqcd\right)\sigma^\LO,\Deltasigmaqaew\right)\right|_\text{PDF}.
  \end{split}
  \label{eq:varpdfgeneral}
\end{equation}
This formula has to be handled with care as we chose to use different PDF sets for different contributions. We can assume that the covariance between the quark--photon-induced channel and every other contribution is independent of the choice of the PDF. This is a valid assumption as the used PDF sets agree reasonably well in their values for quark and gluon PDFs (cf.\ Ref.~\cite{Butterworth:2015oua}) and the covariance is ruled by the quark PDFs. Therefore, Eq.\ \eqref{eq:varpdfgeneral} simplifies to
\begin{equation}
  \begin{split}
    \left.\var\left(\sigma^\NLO\right)\right|_\text{PDF} = &\left.\var\left(\sigma^\NLO\right)\right|_\text{\propername{NNPDF3.0QED}} - \left.\var\left(\Deltasigmaqaew\right)\right|_\text{\propername{NNPDF3.0QED}}\\
    & + \left.\var\left(\Deltasigmaqaew\right)\right|_\text{\propername{LUXqed}}.
  \end{split}
\end{equation}
As the NNPDF collaboration uses Monte Carlo replicas, the variance of a cross section $\sigma$ evaluated with the \propername{NNPDF3.0QED} PDF set is given by
\begin{equation}
  \left.\var(\sigma)\right|_\text{PDF} = \frac{1}{N_\text{rep}-1} \sum\limits_{j=1}^{N_\text{rep}} \left(\sigma_j-\sigma_0\right)^2,
\end{equation}
where $N_\text{rep}$ is the number of replicas, $\sigma_j$ the cross section evaluated with replica $j$ and $\sigma_0=\langle\sigma\rangle$ the central value of the cross section. In contrast to the NNPDF collaboration, the \propername{LUXqed} PDF set uses the Hessian method where the variance is given by the sum over the squared differences between the central value and the contribution evaluated with each eigenvector PDF set (cf.\ Sec.\ \ref{sec:photonpdferror}). 
The results of the PDF uncertainty estimation are shown in Tab.~\ref{tab:nloscalepdf}. Note that using \propername{LUXqed} the impact of the uncertainty of the photon PDF is rather small and negligible in comparison to the scale and other PDF uncertainties. Even using a more conservative approach and taking the \propername{CT14QEDinc} uncertainty would insignificantly change the total PDF uncertainty by only $\sim5\%$ at $13\,\TeV$ for $\PWm\PWp\PWp$ production. 
\begin{table}
  \centering
  \caption{LO and full NLO cross sections with estimated scale (first) and PDF uncertainties (second) at different $\Pp\Pp$ CM energies $\sqrt{s}$.}
  \label{tab:nloscalepdf}
  \begin{subtable}{\linewidth}
    \centering
    \caption{$\Pp\Pp\rightarrow\PWm\PWp\PWp + X$}
    \begin{tabular}{d{3}d{1.1}D{.}{\hphantom{9}}{0.3}d{3.5}d{1.1}D{.}{\hphantom{9}}{0.3}d{3.4}}
      \toprule
      \ccol{$\sqrt{s}$ {\small[TeV]}} & \multicolumn{3}{c}{$\sigma^\LO$ {\small[pb]}} & \multicolumn{3}{c}{$\sigma^\NLO$ {\small[pb]}}\\
      \midrule
      7 & 0.0294 & .^{+0.0009}_{-0.0009} & \pm0.0019 & 0.0442 & .^{+0.0023}_{-0.0019} & \pm0.0014\\
      8 & 0.0371 & .^{+0.0009}_{-0.0009} & \pm0.0023 & 0.0572 & .^{+0.0029}_{-0.0024} & \pm0.0017\\
      13 & 0.0795 & .^{+0.0000}_{-0.0002} & \pm0.0050 & 0.136 & .^{+0.006}_{-0.005} & \pm0.004\\
      14 & 0.0885 & .^{+0.0000}_{-0.0004} & \pm0.0056 & 0.154 & .^{+0.007}_{-0.006} & \pm0.004\\
      100 & 0.98 & .^{+0.09}_{-0.10} & \pm0.07 & 2.657 & .^{+0.004}_{-0.009} & \pm0.055\\
      \bottomrule
    \end{tabular}
  \end{subtable}\\
  \vspace*{1em}
  \begin{subtable}{\linewidth}
    \centering
    \caption{$\Pp\Pp\rightarrow\PWp\PWm\PWm + X$}
    \begin{tabular}{d{3}d{1.1}D{.}{\hphantom{9}}{0.3}d{3.5}d{1.1}D{.}{\hphantom{9}}{0.3}d{3.4}}
      \toprule
      \ccol{$\sqrt{s}$ {\small[TeV]}} & \multicolumn{3}{c}{$\sigma^\LO$ {\small[pb]}} & \multicolumn{3}{c}{$\sigma^\NLO$ {\small[pb]}}\\
      \midrule
      7 & 0.0139 & .^{+0.0004}_{-0.0004} & \pm0.0010 & 0.0216 & .^{+0.0012}_{-0.0010} & \pm0.0009\\
      8 & 0.0181 & .^{+0.0004}_{-0.0004} & \pm0.0012 & 0.0289 & .^{+0.0015}_{-0.0012} & \pm0.0011\\
      13 & 0.0433 & .^{+0.0000}_{-0.0002} & \pm0.0029 & 0.076 & .^{+0.004}_{-0.003} & \pm0.002\\
      14 & 0.0489 & .^{+0.0002}_{-0.0004} & \pm0.0033 & 0.088 & .^{+0.004}_{-0.003} & \pm0.003\\
      100 & 0.72 & .^{+0.07}_{-0.08} & \pm0.05 & 1.999 & .^{+0.006}_{-0.010} & \pm0.048\\
      \bottomrule
    \end{tabular}
  \end{subtable}
\end{table}

We conclude that at past and present LHC energies, the dominant theoretical uncertainty arises from the scale dependence of the NLO prediction, given that modern and up-to-date PDFs are employed. In this case, the scale uncertainties are almost twice as large as the PDF errors and a further improvement on the prediction would require the inclusion of QCD corrections beyond NLO.

\section{Conclusion}
\label{sec:conclusion}

Owing to its sensitivity to the mechanism of electroweak symmetry breaking and to triple and quartic gauge couplings, triple-$\PW$ production is an important process to further test the validity of the SM and search for physics beyond. As precise predictions are necessary to analyze experimental data, we provide full NLO cross sections for the production of three on-shell \PW bosons at proton--proton colliders. We observe that NLO corrections are dominated by QCD with $K$-factors of $\sim1.5{-}1.7$. The electroweak correction are of the order of $\sim5{-}10\%$ at LHC energies. In special kinematic regimes, the electroweak corrections grow large due to high-energy logarithms. The main contribution of the electroweak corrections results from the quark--photon-induced channel, yielding corrections of $\sim11\%$ at $13\,\TeV$. However, we observe large cancellations between the positive corrections from the photon-induced and the negative EW corrections to the quark-induced channels, so that the net EW corrections are at the level of $\sim7\%$. The impact of the quark--photon induced channel can be effectively suppressed by applying a veto on hard jet emissions. We estimate the impact of the uncertainty of the photon PDF on the NLO prediction by considering different PDF sets incorporating QED corrections. Using the recently released LUXqed PDF set, we observe a significant reduction in the uncertainties that originate from the photon PDF and note that the total theory error to this process is now governed by scale uncertainties. To further improve the cross section predictions, it would be necessary to perform at least some multi-jet merging, which is beyond the scope of this work. To improve the predictive power in the distributions one should drop the on-shell requirement and include leptonic decays of the \PW bosons.

\section*{Acknowledgments}
S.D.{ }and G.K.{ }are supported by the Research Training Group GRK 2044 of the German Research Foundation (DFG). S.D.{ }and G.K.{ }acknowledge support by the state of Baden-Württemberg through bwHPC and the German Research Foundation (DFG) through grant no INST 39/963-1 FUGG. A.H.{ }is supported by the Swiss National Science Foundation (SNF) under contract CRSII2-160814.
\appendix
\section{Results at a single phase-space point}

In this appendix, we provide results for the partonic process $\Pqu\Paqd\rightarrow\PWm\PWp\PWp$ at a single phase-space point with the four-momenta given in Tab.~\ref{tab:pspoint}.%
\footnote{
  The process $\Paqu\Pqd\rightarrow\PWp\PWm\PWm$ is trivially obtained by a CP transformation.
}
\begin{table}
  \centering
  \caption{Momenta at a random phase-space point for the process $\Pqu\Paqd\rightarrow\PWm\PWp\PWp$.}
  \label{tab:pspoint}
  \footnotesize
  \begin{tabular}{c cccc}
    \toprule
    particle & $E$ [{\GeV}] & $p_x$ [{\GeV}] & $p_y$ [{\GeV}] & $p_z$ [{\GeV}]\\
    \midrule
    \Pqu  & $159.62609744642299$ & $0$                             & $0$                             & $\phantom{+}159.62609744642299$\\
    \Paqd & $159.62609744642299$ & $0$                             & $0$                             & $-159.62609744642299$\\
    \PWm  & $123.42985984403438$ & $\phantom{+}47.792534737566015$ & $0$                             & $-80.554675217994117$\\
    \PWp  & $112.17654271370813$ & $-43.584978267305615$           & $-14.783517208093180$           & $\phantom{+}63.274211165346756$\\
    \PWp  & $83.645792335103465$ & $-4.2075564702603998$           & $\phantom{+}14.783517208093180$ & $\phantom{+}17.280464052647361$\\
    \bottomrule
  \end{tabular}
\end{table}
The input parameter scheme has been defined in Sec.~\ref{sec:inputparam} and the dynamical scale setting~\eqref{eq:centralscale} reduces to the production threshold, $\mu=3\MW$, for the $2\to3$ kinematics considered here.
In the following, we provide the squared amplitude averaged over initial-state colors and helicities and summed over final-state helicities. 
The virtual corrections are renormalized according to our input-parameter scheme with external legs renormalized on-shell. 
Infrared singularities are regularized using dimensional regularization ($D=4-2\epsilon$) and we further extract a factor of $c_\epsilon=(4\pi)^{\epsilon}\,\Gamma(1+\epsilon)$ from the coefficients of the poles.

At Born level, we obtain
\begin{equation}
  \left\lvert \cM_\mathrm{0} \right\rvert^2 =
  2.1306869301777854 \times 10^{-6} \,\GeV^{-2}
\end{equation}
and the virtual electroweak correction is given by
\begin{equation}
  \begin{aligned}
    2\, \Re\left( \cM_\text{1-loop}^\text{EW} \, \cM_\mathrm{0}^* \right) ={}& -7.0894856389859852 \times 10^{-8} \,\GeV^{-2}\\ 
    &+\frac{c_\epsilon}{\epsilon} \left(-3.8744611130204037 \times 10^{-10} \,\GeV^{-2}\right)\\
    &+\frac{c_\epsilon}{\epsilon^2} \left(-1.4247106236890944 \times 10^{-9} \,\GeV^{-2}\right) .
  \end{aligned}
\end{equation}
For the virtual QCD correction we obtain
\begin{equation}
  \begin{aligned}
    2\, \Re\left( \cM_\text{1-loop}^\text{QCD} \, \cM_\mathrm{0}^* \right) ={}\alphas \Big[&+5.5459644298006651 \times 10^{-6}\,\GeV^{-2}\\
    &+\frac{c_\epsilon}{\epsilon} \left(-8.4905479254227342 \times 10^{-7}\,\GeV^{-2}\right)\\
    &+\frac{c_\epsilon}{\epsilon^2} \left(-9.0429161898421486 \times 10^{-7}\,\GeV^{-2}\right)\Big] ,
  \end{aligned}
\end{equation}
where we have further pulled out a global factor of $\alphas$.

\bibliographystyle{tep}
\bibliography{www}

\end{document}